\documentclass[twocolumn]{aa}  
\usepackage{graphicx}
\usepackage{color}
\usepackage{epsfig}
\usepackage{supertabular}
\usepackage{rotating}
\usepackage{txfonts}
\begin{document}
   \title{The specific frequencies of ultra-compact dwarf galaxies}


   \author{
          S. Mieske
          \inst{1}
          \and
          M. Hilker
          \inst{2}
          \and
          I. Misgeld
          \inst{2}
                   }

   \offprints{S. Mieske}

   \institute{
              European Southern Observatory, Alonso de Cordova 3107, Vitacura, Santiago, Chile
         \and 
              European-Southern Observatory, Karl-Schwarzschild Str. 2, 85748 Garching, Germany
         }

   \date{}

 
\abstract 
{One formation channel discussed for ultra-compact dwarf galaxies
  (UCDs) is that of massive star clusters, and the other main scenario is
  that of tidally transformed dwarf galaxies.}
{We aim at quantifying the specific frequency of UCDs in a range of
  environments and at relating this to the frequency of star clusters
  and potential progenitor dwarf galaxies. Are the frequencies of UCDs
  consistent with being the bright tail of the globular cluster
  luminosity function (GCLF)?}
{We propose a definition for the specific frequency of UCDs,
  $S_{N,UCD} = N_{UCD} 10^{0.4 (M_{V,host}-M_{V,0})} c_{w}$. The
  parameter $M_{V,0}$ is the zeropoint of the definition, chosen such
  that the specific frequency of UCDs is the same as those of globular
  clusters, $S_{N,GC}$, if UCDs follow a simple extrapolation of the
  GCLF. Considering UCDs as compact stellar systems with $M_V<-10.25$
  mag (mass above $\sim 2 \times 10^6$ M$_{\odot}$), it is
  $M_{V,0}=-20$ mag. The parameter $c_{w}$ is a correction term to
  take the dependence of the GCLF width $\sigma$ on the host galaxy
  luminosity into account. We apply our definition of $S_{N,UCD}$ to
  results of spectroscopic UCD searches in the Fornax, Hydra and
  Centaurus galaxy clusters, two Hickson Compact Groups, and the Local
  Group. This includes a large database of 180 confirmed UCDs in
  Fornax. }
{We find that the specific frequencies derived for UCDs match those of
  GCs very well, to within 10-50\%. The ratio
  $\frac{S_{N,UCD}}{S_{N,GC}}$ is $1.00 \pm 0.44$ for the four
  environments Fornax, Hydra, Centaurus, and Local Group, which have
  $S_{N,GC}$ values. This good match also holds for individual giant
  galaxies in Fornax and in the Fornax intracluster-space. The error
  ranges of the derived UCD specific frequencies in the various
  environments then imply that not more than $\sim$50\% of UCDs were
  formed from dwarf galaxies. We show that such a scenario would
  require $\gtrsim$90\% of primordial dwarfs in galaxy cluster centers
  ($<$100 kpc) to have been stripped of their stars.}
{We conclude that the number counts of UCDs are fully consistent with
  them being the bright tail of the GC population. From a statistical
  point of view there is no need to invoke an additional formation
  channel.}

\titlerunning{Specific frequencies of UCDs}


   \maketitle 
%

\section{Introduction}
Ultra-compact dwarf galaxies (UCDs) were recognized as a populous and
potentially distinct class of objects about a decade ago (Drinkwater
et al.~\cite{Drinkw03}), following the results of various
spectroscopic surveys in the Fornax cluster (Minniti et
al.~\cite{Minnit98}, Hilker et al.~\cite{Hilker99}, Drinkwater et
al.~\cite{Drinkw00}, Phillipps et al.~\cite{Philli01}). UCDs are
generally considered as compact stellar systems with masses above
$\simeq$ 2$ \times 10^6$ M$_{\odot}$ and sizes below $\sim$ 100 pc
(e.g. Ha\c{s}egan et al.~\cite{Hasega05}, Mieske et
al.~\cite{Mieske08}). Since their discovery in Fornax, more UCDs have
been detected in a range of environments, from loose and compact
galaxy groups (e.g. Evstigneeva et al.~\cite{Evstig07}, da Rocha et
al.~\cite{Daroch11}) to dense galaxy clusters like Virgo, Centaurus,
Hydra, Coma and Fornax itself (e.g. Jones et al.~\cite{Jones06},
Ha\c{s}egan et al.~\cite{Hasega05}, Misgeld et al.~\cite{Misgel08} \&
~\cite{Misgel11}, Mieske et al.~\cite{Mieske04} \& ~\cite{Mieske07},
Gregg et al.~\cite{Gregg09}, Chiboucas et al.~\cite{Chibou10a} \&
~\cite{Chibou10b}).

The luminosity and size distribution of UCDs shows a smooth transition
to the regime of globular clusters (GCs) (e.g. Ha\c{s}egan et
al.~\cite{Hasega05}, Mieske et al.~\cite{Mieske04} \&
~\cite{Mieske06}). It is thus not surprising that one of the two
discussed formation channels of UCDs is that of massive star clusters
created in the same or a similar way to the main star cluster
population of early-type galaxies (e.g. Fellhauer \& Kroupa
~\cite{Fellha02} \& ~\cite{Fellha05}, Murray ~\cite{Murray09}, Gieles
et al.~\cite{Gieles10}).

The other proposed formation channel is that of tidally stripped dwarf
galaxies (e.g. Zinnecker et al.~\cite{Zinnec88}; Drinkwater et
al. ~\cite{Drinkw03}, Bekki et al.~\cite{Bekki03}, Goerdt et
al.~\cite{Goerdt08}), as has also been frequently suggested for the
Local Group object $\omega$ Cen (e.g. Majewski et al.~\cite{Majews00},
Carraro \& Lia~\cite{Carrar00}, Hilker \& Richtler~\cite{Hilker00},
Bekki \& Freeman~\cite{Bekki03b}, Noyola et al.~\cite{Noyola08}, da
Costa \& Coleman~\cite{daCost08}). Using the mass limit above for the
definition of a UCD, indeed $\omega$ Cen is the only UCD associated with
the Milky Way (see e.g. the Milky Way GC catalog of Harris et
al.~\cite{Harris96}).

In this paper we test whether the numbers of UCDs are consistent with
being merely the bright ($\sim 1\%$) tail of the globular cluster
luminosity function (GCLF). A clear excess of UCDs above the canonical
GCLF would suggest that a separate process, distinct from the GC
formation process, is responsible for the overabundance of UCDs. To
this end we define the specific frequency of UCDs, by relating the
number of UCDs to the luminosity $M_{V,host}$ of their host
galaxy. The zero-point of the definition is chosen such that the UCD
specific frequency is equal to that of GCs for the case
that the numbers of UCDs are equal to a simple extrapolation of the
GCLF. We calculate this specific frequency in a range of environments,
and compare it to that of star clusters and potential progenitor dwarf
galaxies. Special attention is given to the Fornax galaxy cluster,
where a large database of confirmed UCDs and GCs exists.

\section{Definition of specific frequency of UCDs}
\label{sndefsec}

In this section we define the specific frequency $S_{N,UCD}$ of
UCDs. Following e.g. Ha\c{s}egan et al.~(\cite{Hasega05}); Dabringhausen et
al.~(\cite{Dabrin08}); and Mieske et al.~(\cite{Mieske08}) we define UCDs as
compact stellar systems with dynamical masses above $2 \times  10^6$
M$_{\odot}$. To convert this to an absolute magnitude limit, we assume
an optical mass-to-light ratio $M/L_V$ of 2, which is the average
$M/L_V$ of compact stellar systems at this limiting mass (Fig. 12 of Mieske et
al.~\cite{Mieske08})\footnote{For higher masses and metallicities,
  $M/L_V$ increases to larger average values of 4-5 (Mieske et
  al.~\cite{Mieske08})}. This therefore yields an absolute limiting
magnitude between GCs and UCDs of $M_V=-10.25$ mag, assuming that the
Sun's absolute magnitude is $M_{V,\odot}=4.75$ mag.

\begin{figure}
\begin{center}
  \includegraphics[width=8.6cm]{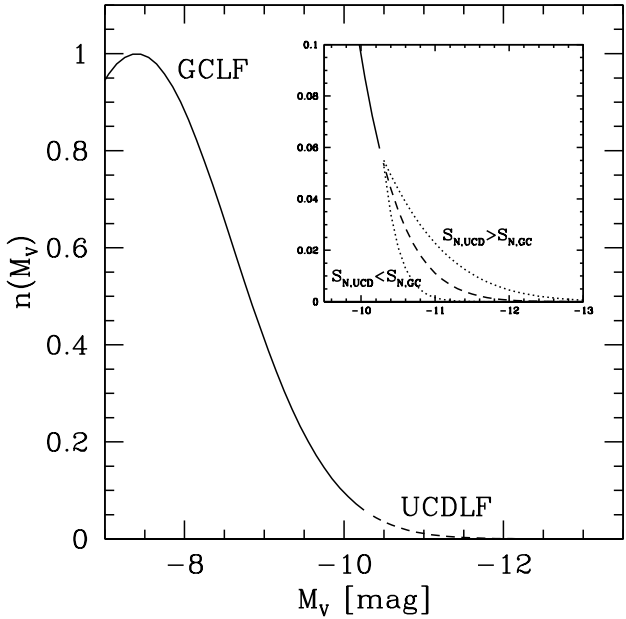}
  \caption{This plot
    illustrates the definition of the UCD specific frequency
    $S_{N,UCD}$. The solid line represents a typical GC
    luminosity function (GCLF) in the shape of a Gaussian, centered at
    $M_V=-7.4$ mag, with a width of $\sigma=$1.2 mag. The line becomes
    dashed in the magnitude range of UCDs $M_V<$ $-10.25$ mag. It holds
    that $S_{N,UCD}=S_{N,GC}$ if the luminosity distribution of UCDs
    follows the extrapolated GCLFs, and that $S_{N,UCD} > S_{N,GC}$ if
    the frequency of UCDs is above the GCLF extrapolation (upper
    dotted line in inset), and $S_{N,UCD} < S_{N,GC}$ if it is below
    (lower dotted line in inset).}
\label{sndefplot}
\end{center}
\end{figure}

One formation scenario of UCDs is that they constitute the high-mass
tail of the star cluster population (e.g. Hilker et
al.~\cite{Hilker99}, Drinkwater et al.~\cite{Drinkw00}, Fellhauer \&
Kroupa \cite{Fellha02}, Mieske et al.~\cite{Mieske02} \&
\cite{Mieske04}, Murray \cite{Murray09}, Gieles et
al.~\cite{Gieles10}, Frank et al.~\cite{Frank11}). Therefore, our aim
is a quantitative comparison of the specific frequency of UCDs to
those of GCs. To calculate the UCD specific frequency we
relate the number of UCDs to the luminosity $M_{V,host}$ of its host
galaxy, analogous to the case of GCs (Harris \& van den
Bergh~\cite{Harris81}). The prerequisite in that context is that the
specific frequency of UCDs should be equal to that of GCs if the
luminosity distribution of UCDs is consistent with a simple
extrapolation of the GCLF to brighter magnitudes. This is illustrated
in Fig.~\ref{sndefplot} for the default assumption of a Gaussian GCLF
$N(M_V) \propto e^{\frac{-(M_{V,TOM}-M_V)^2}{2 \sigma^2}}$.

The term $M_{V,TOM}$ denotes the turnover-magnitude (=TOM) of the
GCLF, which is the maximum of the log-normal luminosity function. The
term $\sigma$ denotes the characteristic width of that GCLF, which has
typical values between 1.0 and 1.4 mag. For the UCD specific frequency
we choose the same functional form that is used for the specific
frequency of GCs (Harris \& van den
Bergh~\cite{Harris81}; Peng et al.~\cite{Peng08} and references
therein).

\begin{equation}
S_{N,UCD} = N_{UCD}   10^{0.4   (M_{V,host}-M_{V,0})}   c_{w}
\label{sndef}
\end{equation}

For $c_{w}=1$, $M_{V,0}$ is the galaxy luminosity at which for a
specific frequency $S_{N,UCD}=1$ one would expect exactly one
UCD. Given the premise to make $S_{N,UCD}$ directly comparable to
$S_{N,GC}$, the value of $M_{V,0}$ depends on which fraction of the
area below the canonical GCLF is occupied by the UCD luminosity
range. The correction term $c_{w}$ is included to account for the
varying width of the GCLF as a function of host galaxy magnitude
(e.g. Jord\'{a}n et al.~\cite{Jordan07}). If the GCLF of the
investigated host galaxy is identical to the reference GCLF (see
below), $c_w=1$. The exact functional form of $c_w$ as a function of
GCLF width is discussed at the end of this Section. For convenience in
the further course of the paper, we also introduce the `reference'
specific frequency $S_{N,UCD}^*=S_{N,UCD} / c_w$, which is the
frequency for when all galaxies have the same (reference)
GCLF width.

We furthermore define the quantity $n_{UCD}$ as the fraction of
sources below a normalized Gaussian GCLF that have luminosities
$M_V<-10.25$ mag, the UCD luminosity limit. For $M_{V,0}$ it then holds that

\begin{equation}
M_{V,0}=-15 + 2.5   \log{n_{UCD}}  .
\end{equation}

The term $-15$ comes directly from the definition of the GC specific
frequency, for which $M_{V,0}=-15$ mag (Harris \& van den
Bergh~\cite{Harris81}). In the following we consider a Gaussian GCLF
with a turnover-magnitude of $M_{V,TOM}=-7.4$ mag
(e.g. Harris~\cite{Harris91}, Kundu \& Whitmore~\cite{Kundu01}, Jord\'{a}n
et al.~\cite{Jordan06}, Peng et al.~\cite{Peng08}), see also
Fig.~\ref{sndefplot}. The width of the GCLF is typically $\sigma =
1.2$ mag for galaxies with absolute magnitudes of around $M_V=-22$
mag. As reference width we adopt $\sigma=1.223$ mag, since this
corresponds exactly to the case where 1\% of the area below the GCLF
falls into the UCD magnitude range $M_V<-10.25$ mag, hence
$n_{UCD}=0.01$. With this reference we obtain a `simple' value of
$M_{V,0}=-20$ mag, five magnitudes offset from the zeropoint of the GC
specific frequency. For a UCD specific frequency of unity in the case
of the Milky Way ($M_V=-20.5$ mag), one would thus expect 1-2 UCDs,
which corresponds to reality since $\omega$ Cen is the only MW
satellite that satisfies the UCD definition. At the same time, a
specific frequency around unity is also typically found for the GC
populations of Milky Way type galaxies (e.g. Goudfrooij et
al.~\cite{Goudfr03}; Rhode \& Zepf~\cite{Rhode04}; Chandar et
al.~\cite{Chanda04}). Given the prerequisites we imposed for the
definition of $S_{N,UCD}$, this shows that the number of UCDs in the
Milky Way is about consistent with what is expected from an extrapolation
of its GCLF. In the next section this is discussed in more
detail for a range of environments.

While for the Milky Way and Local Group the luminosity distribution of
compact stellar systems is sampled down to very faint absolute
magnitudes well into the regime of bona-fide GCs, this
is not necessarily the case for most galaxy clusters where UCDs have
been found (Drinkwater et al.~\cite{Drinkw00}, Phillipps et
al.~\cite{Philli01}, Mieske et al.~\cite{Mieske04} \&~\cite{Mieske07},
Misgeld et al.~\cite{Misgel11}), at distance moduli in the range
$31<(m-M)<33.5$ mag, and even 35 mag for Coma (Chiboucas et
al.~\cite{Chibou10a}~\&~\cite{Chibou10b}). The completeness of the
surveys within $(m-M)<33.5$ mag in terms of slit allocation, area
coverage, and spectroscopic success rate is well defined and around
50\% for $M_V \lesssim -11$ mag. However, at fainter magnitudes the
available survey data are very heterogeneous in terms of target
selection, slit allocation, area selection, and spectroscopic 'success
rate' (see references above). This needs to be accounted for in the
calculation of the specific frequency of UCDs. We therefore define an
equivalent formulation of $S_{N,UCD}$ for the case that the UCD sample
is only well known for $M_V < -11$ mag. In the context of the
$S_{N,UCD}$ definition above, the change we need to adopt for 
restricting the UCD sample to $M_V<-11$ mag is the value of
$M_{V,0}$. The limit of $M_V<-11$ mag corresponds to 2.95$\sigma$ away
from the turnover magnitude for a reference GCLF width $\sigma=1.223$
mag, yielding a fraction of 0.00159 of all sources. This is 6.29 times
less than for the default limit of $M_V<-10.25$ mag, or quite exactly
2.0 magnitudes. Therefore, we adopt $M_{V,0}=-22$ mag when the UCD
sample is restricted to $M_V<-11$ mag. The two formulations with
different $M_{V,0}$ are considered equivalent in the context of our
study, where we want to test the consistency of the UCD luminosity
distribution with the extrapolation of the GCLF.

\subsection{The correction term $c_w$}

The fractional area $A$ below a Gaussian GCLF brighter than a given
magnitude $M_V$ is related to the error function $erf$ in the following
way

\begin{equation}
A = 1-0.5 (1+erf(\frac{M_V+7.4}{\sigma   \sqrt{2}}))  .
\end{equation}

\noindent From this and the reference of $\sigma=1.223$, it follows for the
definition of $c_w$ for a fully sampled UCDLF down to
$M_V<-10.25$ mag that

\begin{equation}
c_w=\frac{1-0.5 (1+erf(\frac{2.85}{\sigma  
    \sqrt{2}}))}{1-0.5 (1+erf(\frac{2.85}{1.223  
    \sqrt{2}}))}   .
\label{cwall}
\end{equation}

\noindent For a partially sampled UCDLF down to $M_V<-11$ mag the following
then holds:

\begin{equation}
c_w=\frac{1-0.5 (1+erf(\frac{3.6}{\sigma  
    \sqrt{2}}))}{1-0.5 (1+erf(\frac{3.6}{1.223  
    \sqrt{2}}))}
\label{cw11}
\end{equation}

It is well known that the width $\sigma$ of the GCLF depends on the
host galaxy luminosity $M_{V,host,individual}$, becoming higher for
brighter host galaxies (e.g. Jord\'{a}n et al.~\cite{Jordan07}). To take
this into account for the correct determination of $c_w$, we adapt
equation (18) of Jord\'{a}n et al. (~\cite{Jordan07}) to calculate
$\sigma$ as a function of $M_{V,host,individual}$:

\begin{equation}
\sigma = 1.14 - 0.100   (M_{V,host,individual}+20.9)  .
\label{sigdef}
\end{equation}

\noindent For this we have explicitly assumed (B-V)=0.9 mag, which is
the color expected for a $\sim$10 Gyr single-burst stellar population
of [Fe/H] $\sim -$0.7 dex (Bruzual \& Charlot ~\cite{Bruzua03}). We
furthermore assume that the GCLF width in the SDSS g-band (Jord\'{a}n et
al.~\cite{Jordan07}) is identical to the width in the V-band. It is
important to note that the term $M_{V,host,individual}$ refers to the
luminosity of an individual host galaxy. For the calculation of the
specific frequency, we need to normalize the number of UCDs to the total
galaxy luminosity in the surveyed area, which can be some 0.3-0.5 mag
brighter than the luminosity of the brightest individual galaxy (see
next section).

\section{Specific frequency as a function of environment}
\label{envsn}

\begin{figure*}
\begin{center}
\includegraphics[width=8.6cm]{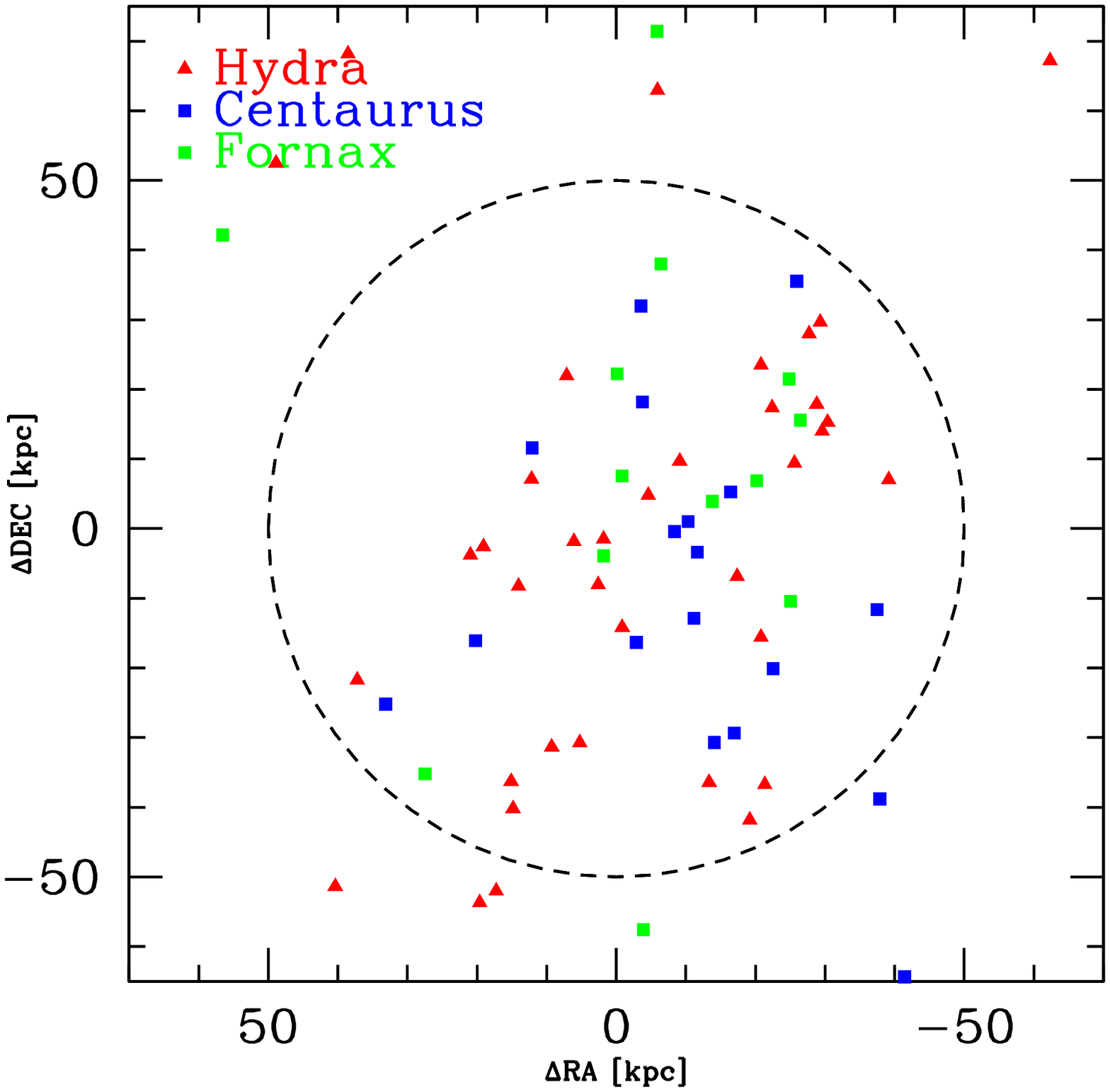}
\includegraphics[width=8.6cm]{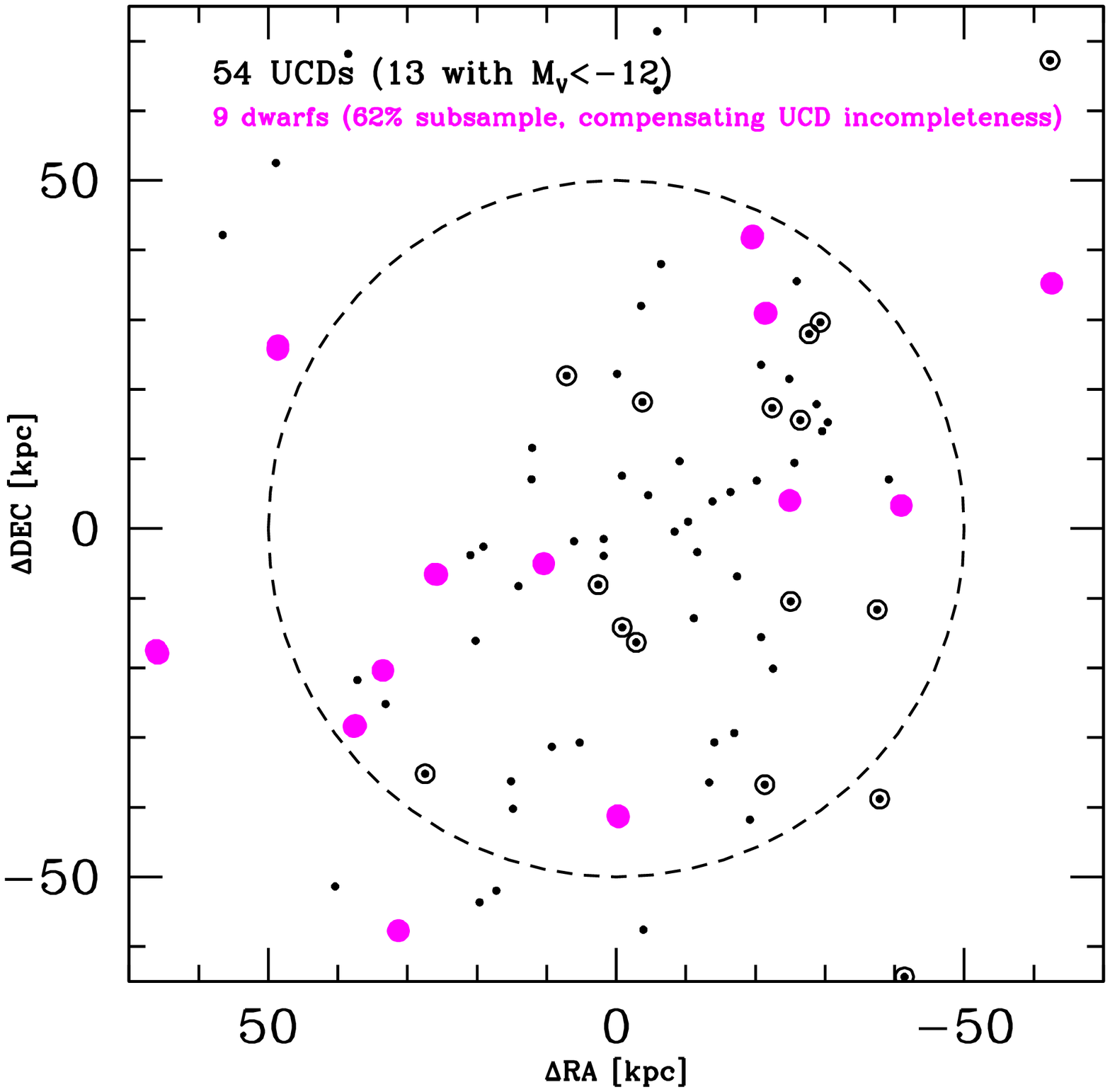}\\

  \caption{Projected spatial distribution of UCDs within the
    inner $\simeq$ 100 kpc of the Hydra, Centaurus and Fornax clusters
    (Misgeld et al.~\cite{Misgel09}, \cite{Misgel11}; Mieske et
    al. \cite{Mieske04}). UCDs are in this context defined as compact
    stellar systems with $M_V<-11$ mag. {\bf Left panel:} UCDs color
    coded according to host cluster. Red triangles are Hydra UCDs, blue squares are Centaurus UCDs, and green circles are Fornax UCDs. {\bf Right panel:} The same
    sample of UCDs is shown as black dots, with the brightest UCDs
    ($M_V<-12$ mag) marked by large open circles. There are 54 UCDs
    known within the central 50 kpc of Hydra, Centaurus, and
    Fornax. Large filled circles indicate the projected positions of
    canonical dwarf galaxies (see text for references), restricted to
    a magnitude range $-20.5<M_V<-16$ mag, which corresponds to the
    assumed luminosity range for possible UCD progenitor galaxies (see
    text for details). To allow a direct comparison of the number
    counts between UCDs and progenitors, we show only 62\% of the
    dwarf galaxies, given that the UCD searches are on average
    complete to $\sim$62\% in the central 50 kpc radius (see text for
    details). }
\label{distribution}
\end{center}
\end{figure*}

In this section we calculate the specific frequency of UCDs in the
three massive nearby galaxy clusters Fornax, Centaurus, and Hydra,
based mainly on data of spectroscopic surveys performed within our
group, except for Fornax for which a wealth of datasets is
available. The projected spatial distribution of UCDs in those
surveys is shown in Fig.~\ref{distribution}. We do not include the
recent data from the Coma cluster (Chiboucas et al.~\cite{Chibou10a})
due to its greater distance, larger incompleteness in terms of
magnitude coverage, and complex selection function of UCD
candidates. Furthermore we calculate the UCD specific frequency
for the Local Group, and two compact galaxy groups recently
investigated in Da Rocha et al. (\cite{Daroch11}). The results of
these calculations are summarized in Table~\ref{tablesn} and
Fig.~\ref{sn}, in which we also compare the derived UCD specific
frequencies with the frequency of GCs.

\subsection{Fornax}

In this section we analyze the specific frequencies of UCDs in
  the Fornax cluster. We start with an analysis of the very central
  Fornax part, based on our UCD survey from Mieske et
  al.~(\cite{Mieske04}), for which the survey completeness is 
  quantified well and which extends down to $M_V\simeq -10.4$ mag. Then we
  extend the analysis to the entire cluster by including all available
  literature results. This includes also the earlier
    Fornax survey performed with the 2dF spectrograph (e.g. Drinkwater
    et al. ~\cite{Drinkw00}), which had a complete areal coverage in the
    central cluster, but a significantly brighter magnitude limit $M_V
    \lesssim -12$ mag. Assumed distance modulus of Fornax is 31.4
    mag (Ferrarese et al.~\cite{Ferrar00}). 

\subsubsection{The central region}

In Mieske et al. (\cite{Mieske04}) we describe our search for
compact stellar systems in the central $\sim$100 kpc radius ($\sim$20
arcminutes) of the Fornax cluster, which was performed with the WFCCD
camera at the 2.5m du Pont telescope at Las Campanas Observatory. In
Fig.~\ref{distribution} we show the location of spectroscopically
confirmed UCDs with $M_V<-11$ mag, restricted to the inner 70-80 kpc
around NGC 1399. Some of these UCDs are outside of the plot. In
Mieske et al.  (\cite{Mieske04}), we calculated the overall survey
completeness for $M_V<-11$ mag to be 70\%$\pm 5\%$ in our survey area,
which comprises the two bright gE galaxies NGC 1399 and NGC 1404. This number takes the varying radial density of
confirmed cluster members compared to background sources into account. The
completeness in terms of slit allocation and spectroscopic success is
calculated in rings, and the global value of the completeness is the
{\it weighted mean} of those values, weighted by the number of UCDs
found in each ring.

There are 20 confirmed UCDs with $M_V<-11$ mag in our survey. Given
the global completeness of 70 \%, this translates into a total number
of 28.6 $\pm$ 7 UCDs with $M_V<-11$ mag in the central Fornax
cluster. In the surveyed region the two gE galaxies NGC 1399 and NGC
1404 contribute the lions share of the galaxy light. NGC 1399 has
$M_V=-22.6$ mag within $\sim$10$'$ radius ($\simeq$ 55 kpc; Dirsch et
al.~\cite{Dirsch03}), while NGC 1404 is about 1 mag fainter (NED). The
sum of both absolute magnitudes therefore corresponds to $M_V \simeq
-23.0$ mag. We assume an uncertainty of 0.2 mag for this absolute
magnitude based on experience with light profile fitting and the
related uncertainty in the accurate determination of the sky
background. We thus obtain a reference specific frequency of
$S_{N,UCD}^*$=11.4 $\pm$ 4 for the central 100 kpc of Fornax. If we
restrict this calculation to the inner $\sim$10$'$ (=55 kpc), we can
disregard the contribution of NGC 1404, but at the same time only have
12 UCDs with $M_V<-11$ mag. This would yield a lower specific
frequency of $S_{N,UCD}^* \simeq$ 7. We adopt the mean of both
estimates $S_{N,UCD}* \simeq 9 \pm 3$. Evaluating $c_w$ for
$M_V=-22.6$ mag (the luminosity of the gE NGC 1399, which dominates in
terms of associated GCs) then yields a final value of $S_{N,UCD} = 4.9
\pm 1.7$. This downward correction is because the GCLF width $\sigma$
for $M_V=-22.6$ mag is 1.31 mag according to Eq.~\ref{sigdef}, which
implies almost twice as many sources in the bright end tail $M_V<-11$
mag than for the reference width of $\sigma=1.223$ mag. For
comparison, the GC specific frequency of NGC 1399 is $S_{N,GC} = 5.1
\pm 1.2$ (Dirsch et al. 2003), fully consistent with the value for
UCDs.

\begin{figure*}
\begin{center}
\includegraphics[width=18.2cm]{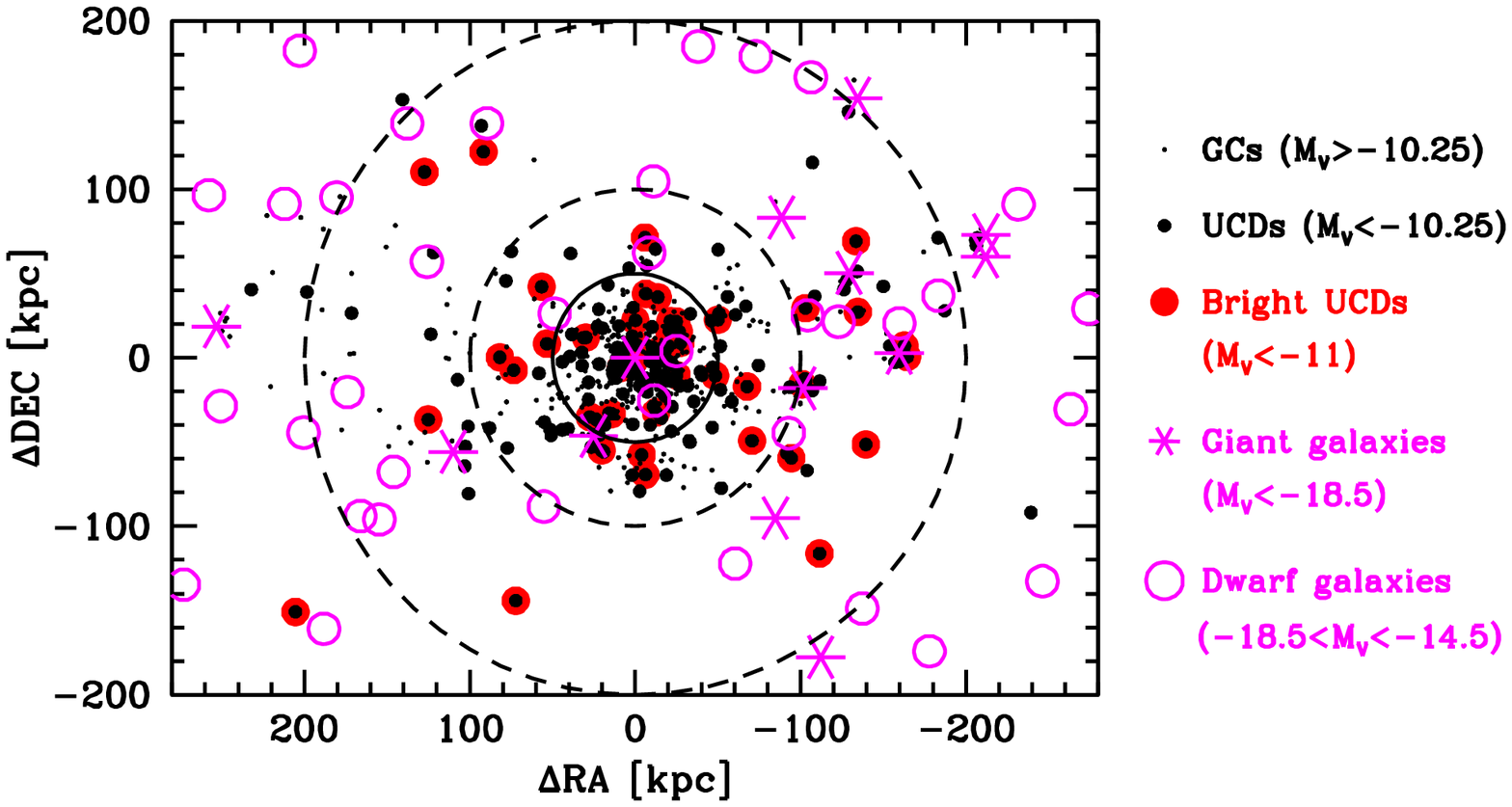}

  \caption{Projected distribution of all confirmed UCDs and GCs in the
    Fornax cluster (see text; spectroscopy and photometry compiled
    from Kissler-Patig et al.~\cite{Kissle99}, Drinkwater et
    al.~\cite{Drinkw00}, Dirsch et al.~\cite{Dirsch03}, Mieske et
    al.~\cite{Mieske04}, Bassino et al.~\cite{Bassin06}, Bergond et
    al.~\cite{Bergon07}, Firth et al.~\cite{Firth07} \&
    \cite{Firth08}, Jord\'{a}n et al.~\cite{Jordan07}, Schuberth et
    al.~\cite{Schube09}, Gregg et al.~\cite{Gregg09}, Puzia et
    al. 2011 private communication). Small black dots are GCs
    ($M_V>-10.25$ mag; about 400 sources). Filled black circles are
    UCDs ($M_V<-10.25$ mag; about 180 sources). Filled red circles are
    `bright' UCDs with $M_V<-11$ mag, about 45 sources. Large magenta
    asterisks indicate Fornax cluster giant galaxies from the FCC,
    defined as having $M_V<-18.5$ mag. Magenta open circles indicate
    dwarf galaxies from the FCC with $-18.5<M_V<-14$ mag, the
    postulated approximate magnitude range of UCD progenitor galaxies
    (Bekki et al.~\cite{Bekki03}). The inner solid circle indicates a
    radius of 50 kpc adopted in Fig.~\ref{distribution} for the
    comparison between the Fornax, Hydra, and Centaurus
    environments. As can be seen, the area coverage of the available
    Fornax spectroscopic surveys drops strongly beyond 50 kpc radius.}
\label{fornax_map}
\end{center}
\end{figure*}
\subsubsection{The overall cluster}

Restricting our considerations in the previous section to the very
central Fornax cluster with only one survey has the advantage of a
homogeneous catalog and of well controlled completeness. However, for
the Fornax cluster there is an extraordinarily large database of
further spectroscopcically confirmed compact cluster members, which is
worth analyzing in the context of the present study. The caveat is
that this database is heterogeneous in survey design and coverage.

\begin{figure}
\begin{center}
\includegraphics[width=8.6cm]{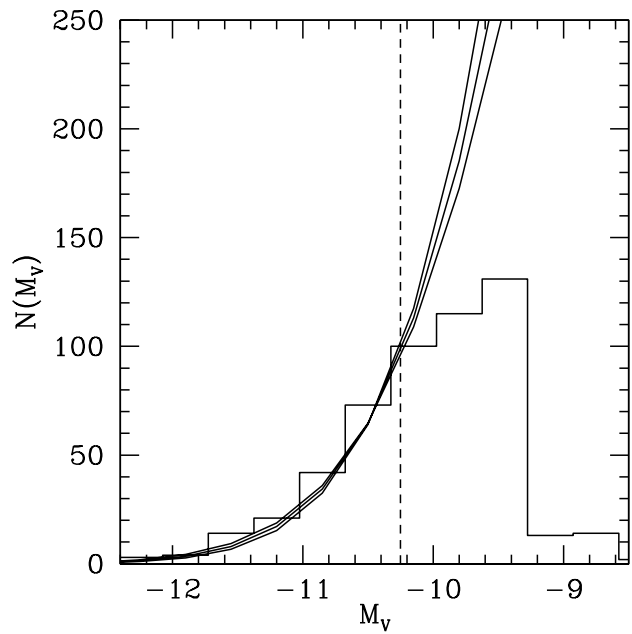}
\caption{Absolute magnitude distribution of GCs and UCDs shown in Fig.~\ref{fornax_map}. The dotted vertical line indicates the magnitude limit
    adopted between UCDs and GCs. The curves show Gaussian GCLFs
    normalized to match the object counts at the UCD magnitude
    limit. The three lines correspond to different widths $\sigma$ of
    the Gaussians: 1.30, 1.35, and 1.40 mag. The completeness of the
    spectroscopic surveys drops at about the UCD limit.}
\label{fornax_magdist}
\end{center}
\end{figure}

In Fig.~\ref{fornax_map} we show the distribution of all compact
objects (UCDs + GCs) in the Fornax cluster (adopted from Fig.1 of
Hilker~\cite{Hilker11}). Their magnitude distribution is shown in
Fig.~\ref{fornax_magdist}. The spectroscopy and photometry has been
compiled from Kissler-Patig et al.~(\cite{Kissle99}), Drinkwater et
al.~(\cite{Drinkw00}), Dirsch et al.~(\cite{Dirsch03}), Mieske et
al.~(\cite{Mieske04}), Bassino et al.~(\cite{Bassin06}), Bergond et
al.~(\cite{Bergon07}), Firth et al.~(\cite{Firth07} \& \cite{Firth08}),
Jord\'{a}n et al.~(\cite{Jordan07}), Schuberth et al.~(\cite{Schube09}),
Gregg et al.~(\cite{Gregg09}), Puzia et al. (2011 private
communication). Small black dots are GCs ($M_V>-10.25$ mag; about 400
sources). Filled black circles are UCDs ($M_V<-10.25$ mag; about 180 sources).
Filled red circles are `bright' UCDs with
$M_V<-11$ mag, about 45 sources. Large magenta asterisks indicate
Fornax cluster giant galaxies from the FCC (Fornax Cluster Catalogue;
Ferguson \& Sandage~\cite{Fergus88}), defined as having $M_V<-18.5$
mag. Magenta open circles indicate dwarf galaxies from the FCC with
$-18.5<M_V<-14$ mag, the postulated approximate magnitude range of UCD
progenitor galaxies (Bekki et al.~\cite{Bekki03})\footnote{See also
  discussion on the progenitor magnitude range in
  Sect.~\ref{progen}}. The inner solid circle indicates the radius of
50 kpc, which is adopted for the comparative analysis of the Fornax,
Hydra and Centaurus environments (see previous and next sections).

The area coverage of the available Fornax spectroscopic surveys starts
to become patchy beyond 50 kpc. Many detections between 100 and 200
kpc clustercentric distance are from the VLT/FLAMES survey of Bergond
et al.~(\cite{Bergon07}), which focuses on a strip of 500 $\times$ 150
kpc aligned along the east-west axis to cover the giant galaxy
distribution in the central Fornax cluster. Another important
contribution comes from the VLT/FLAMES survey of Firth et
al.~(\cite{Firth07}) in the inner 130 kpc radius of Fornax. These
authors quote a spatial completeness of 15-30\% between 80 and 180 kpc
radius. Overall, from these literature sources and
Fig.~\ref{fornax_map} we estimate a spatial coverage of 60-70\%
between 50 and 100 kpc and 30-50\% beyond 100 kpc.

\vspace{0.1cm}

\begin{figure*}
\begin{center}
\includegraphics[width=8.6cm]{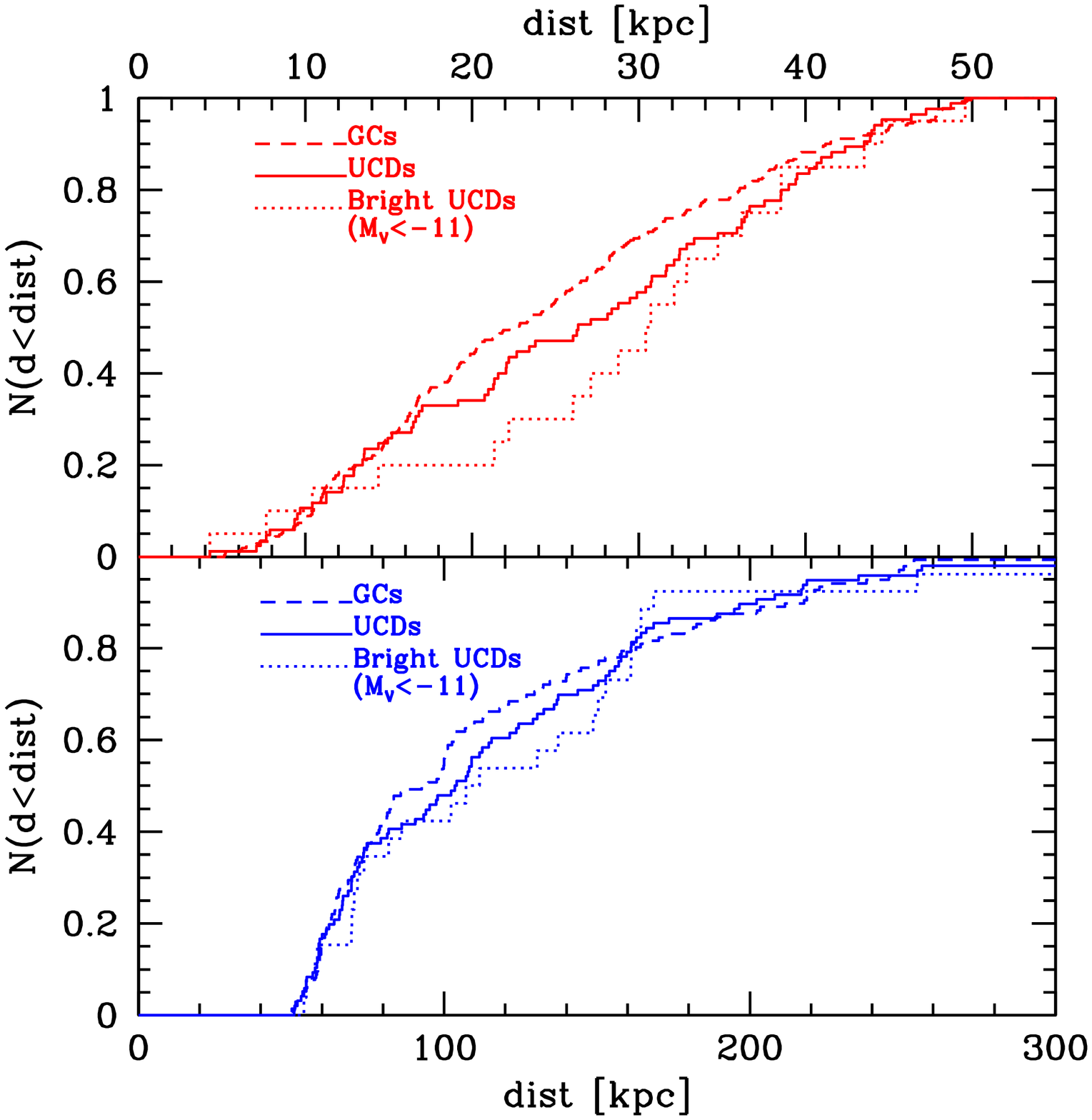}
\includegraphics[width=8.6cm]{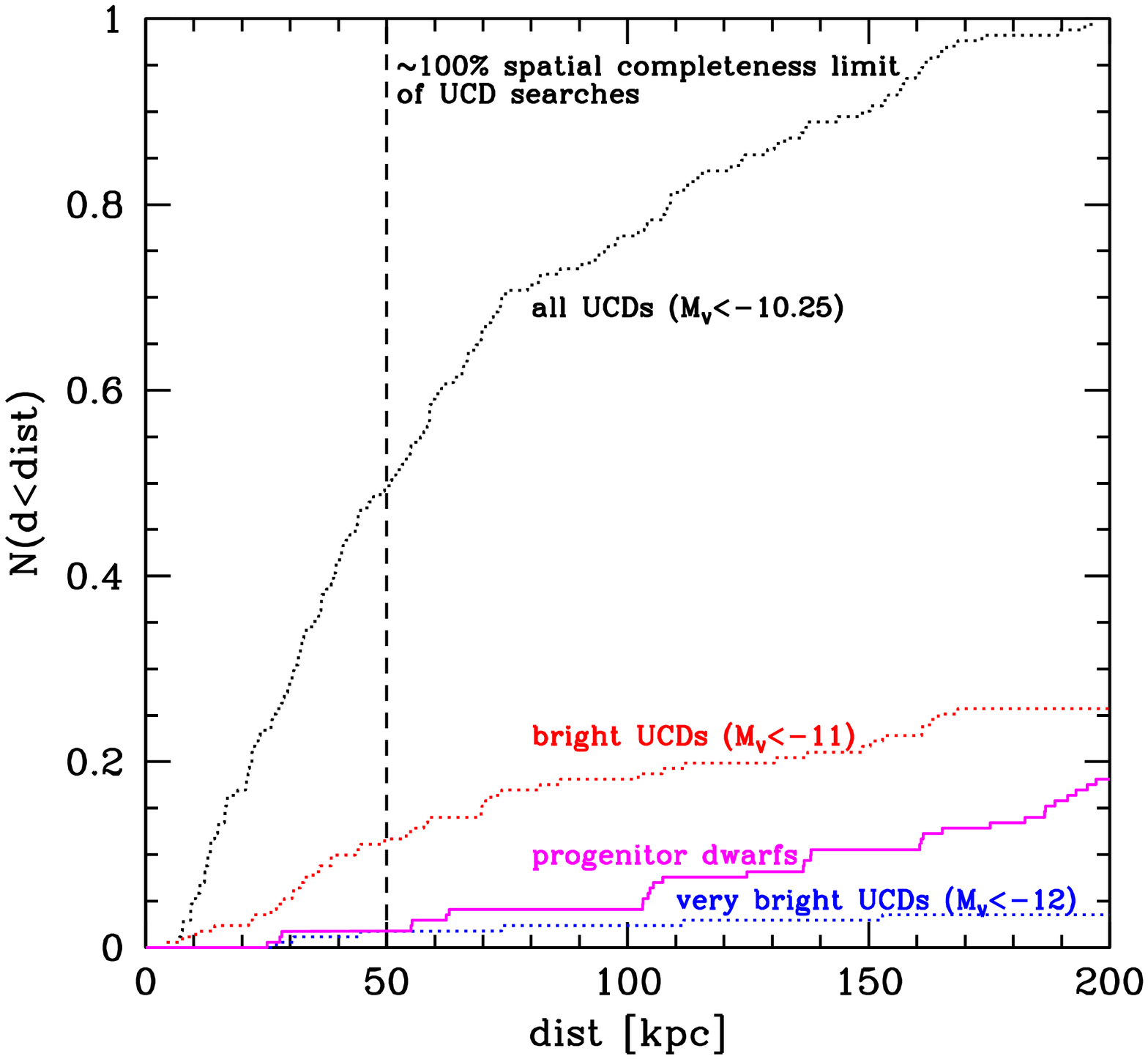}

  \caption{{\bf Left panel:} Projected radial distribution of GCs \&
    UCDs in the Fornax cluster, see also Fig.~\ref{fornax_map}. The
    upper plot with the red curves corresponds to the distribution
    within 50 kpc, with x-axis limits on the top. The lower plot
    with the blue curves corresponds to the distribution outside of 50
    kpc, with x-axis limits on the bottom. Solid lines are UCDs
    ($M_V<-10.25$ mag). Dotted lines are bright UCDs with $M_V<-11.0$
    mag. Dashed lines are GCs ($M_V>-10.25$ mag). There is mild
    evidence (92$\%$ significance according to a KS test) of a
    more extended distribution of bright UCDs compared to GCs within
    the inner 50 kpc. {\bf Right panel:} Cumulative radial
    distribution of UCDs (black, red, and blue curves), compared to
    that of dwarf galaxies (magenta) in the postulated progenitor
    magnitude range of $-18.5<M_V<-14$ mag (Bekki et
    al. ~\cite{Bekki03}); see also Fig.~\ref{fornax_map}. The maximum
    of the various curves have been re-normalized to represent the
    respective total number of objects in each sample. }
\label{fornax_rad}
\end{center}
\end{figure*}

\noindent Based on the data from Fig.~\ref{fornax_map} we address four aspects
that go beyond the focus of the previous subsection:

\vspace{0.1cm}

\begin{enumerate}

\item {\bf What is the specific frequency of UCDs in the inner 50 kpc
  around NGC 1399, including all literature detections down to
  $M_V<-10.25$ mag?} The data shown in Fig.~\ref{fornax_map} give a
  total number of 84 spectroscopically confirmed UCDs with
  $M_V<-10.25$ mag, and 20 UCDs with $M_V<-11$ mag in the inner 50
  kpc. The latter number is consistent with the completeness
  correction of the survey in Mieske et al.~(\cite{Mieske04}), which
  would predict $\sim$17 UCDs with $M_V<-11$ mag in the inner 50
  kpc. Also, the magnitude distribution of confirmed UCDs \&
  GCs (Fig.~\ref{fornax_magdist}) is fully consistent with the GCLF
  shape for $M_V\lesssim -10.5$ mag, while the spectroscopic
  incompleteness becomes notable for $M_V\gtrsim -10$ mag. For the
  innermost 50 kpc, we therefore use the working hypothesis of a
  complete sample down to the UCD limit of $M_V=-10.25$ mag.

We thus find $S_{N,UCD}=6.2 \pm 2.4$ for considering UCDs with
$M_V<-11$ mag, and $S_{N,UCD}=4.1 \pm 0.5$ for considering UCDs with
$M_V<-10.25$ mag. Both values agree with the result derived from the
UCD sample of Mieske et al.~(\cite{Mieske04}) alone and with the specific
frequency of GCs in NGC 1399 (Dirsch et al. ~\cite{Dirsch03}).

\vspace{0.1cm}

\item {\bf How does the number of UCDs associated to individual
  galaxies outside of 50 kpc compare to the prediction assuming
  $S_{N,UCD}=3$?} The particular choice of $S_{N,UCD}=3$ is to
  represent typical GC specific frequencies in the Fornax cluster
  giant galaxies, which range from 1.5 to 5 (Dirsch et
  al.~\cite{Dirsch03}, Bassino et al.~\cite{Bassin06b}, Kissler-Patig
  et al.~\cite{Kissle07}). Integrating for $S_{N,UCD}=3$ over all nine
  giant galaxies beyond NGC1399/1404 which fall into UCD survey
  regions, one would expect a total of $\sim$18 UCDs associated to
  those galaxies. To test this, we restricted to UCDs within 20 kpc
  projected distance from the galaxies. For the Milky Way, this
  restriction would encompass 85-90\% of GCs
  (Harris~\cite{Harris96}). From the data in Fig.~\ref{fornax_map} we
  then find a total of 25 UCDs associated to the nine giant
  galaxies. This number is consistent with the 18 UCDs predicted
  for $S_{N,UCD}=3$. Given the possible incompleteness in the target
  selection and spectroscopic success rates of the various surveys,
  these are lower limits to the number of existing UCDs.

\vspace{0.1cm}

\item {\bf How does the number of spectroscopically confirmed UCDs in
  the intracluster region (100-200 kpc distance) compare to the
  GC surface density determined statistically from
  photometry (Bassino et al.~\cite{Bassin06})?}  From
  Fig.~\ref{fornax_map} it is clear that many UCDs and GCs are found
  in regions that are not directly associated to any giant galaxy.
  Between a clustercentric radius of 100 and 200 kpc we find a total
  of 40 UCDs with $M_V<-10.25$ mag, and about ten UCDs with $M_V<-11$
  mag. Due to the low number counts we focus on UCDs with
  $M_V<-10.25$ mag. Of those, 25 are associated to giant galaxies (see
  previous item). We are thus left with a sample of $\sim$15$\pm$5
  UCDs in the intracluster space between 100 and 200 kpc. In the ring
  between 100 kpc and 200 kpc, we expect about 1000-1100 GCs according
  to the photometrically estimated surface density of GCs in
  that region measured by Bassino et al.~(\cite{Bassin06}), Table
  2. For a GCLF width of $\sigma=1.31$ mag for NGC 1399 (see previous
  subsection), we would thus expect $\sim$20 sources in the UCD
  magnitude regime $M_V<-10.25$ mag. This matches the confirmed
  number of $\sim$15 intracluster UCDs well. However the
  latter value is a lower limit to the true number of UCDs given that
  only about one third of the area between 100 kpc  and 200 kpc has been surveyed
  spectroscopically (Fig.~\ref{fornax_map} and text above). A complete
  spectroscopic survey in this region may still lead to an
  overabundance of UCDs with respect to the IC GCLF of up to a factor
  of $\sim$2.

\vspace{0.1cm}

\item {\bf Is there any evidence for different spatial
  distributions of UCDs and GCs?} In Fig.~\ref{fornax_rad} we show the
projected radial distribution of spectroscopically confirmed UCDs and
GCs from Fig.~\ref{fornax_map}. GCs are shown with the dashed
lines. UCDs ($M_V<-10.25$ mag) are shown as solid lines, and bright
UCDs ($M_V<-11$ mag) as dotted lines. The red curves correspond to the
samples restricted to the inner 50 kpc. The blue curves correspond to
the sample excluding the inner 50 kpc. No significant distribution
difference between GCs and UCDs is seen for the outer region beyond 50
kpc. For the inner region within 50 kpc, there is mild evidence of a
more extended distribution of bright UCDs ($M_V<-11$ mag) compared to
GCs. The KS test shows an 8\% probability that both samples are drawn
from the same parent distribution. This corroborates a similar finding
by Mieske et al.~(\cite{Mieske04}), which shows a 12\% probability
that sources with $M_V<-11$ mag have the same parent distribution as
sources with $M_V>-11$ mag (see also Fig.2 of Hilker~\cite{Hilker11}). 

\end{enumerate}

We conclude that the specific frequencies of confirmed UCDs in the
Fornax cluster agree with that of GCs to within their
errors. Also a comparison of spatial distribution shows only mild
differences between UCDs and GCs. 

The specific frequencies of UCDs will clearly rise to some extent,
once complete spectroscopic surveys down to $M_V \sim -10$ mag are
available for the central Mpc of the Fornax cluster. It is, however,
difficult to predict the magnitude of this effect, or to predict whether it will
give a significant difference between $S_{N,UCD}$ and
$S_{N,GC}$. Based on the rectangular survey coverage
(Fig.~\ref{fornax_map}, Bergond et al.~\cite{Bergon07}) one may
speculate that the number of UCDs could still rise by up to a factor
of $\sim$2 in the outer cluster parts. However, this depends on the
actual spatial distribution of UCDs in these unsurveyed regions, which
is unknown.  The overall galaxy distribution is clearly elongated
along the survey coverage of Bergond et al.~(\cite{Bergon07}), such
that a factor of two increase in UCD number counts from the uncovered
regions appears a reasonable upper limit.

\subsection{Centaurus}

For calculating the specific frequencies of UCDs in the Centaurus
cluster, we used the data presented in Mieske et
al. (\cite{Mieske07},~\cite{Mieske09}), based on two surveys performed
with VIMOS@VLT. Figure~\ref{distribution} shows a map of the central
Centaurus cluster regions with the same physical plot limits as for
Fornax. The assumed distance modulus of Centaurus is 33.28 mag (Mieske
et al. ~\cite{Mieske05}). About three quarters of the UCDs in
Centaurus were detected in one central VIMOS quadrant around the main
galaxy NGC 4696, whose dimension is 7$\times $7 $'$, or 85$\times $85
kpc (see e.g. Fig.1 of Mieske et al.~\cite{Mieske07}). We therefore
restrict our considerations to this pointing and central quadrant and
assume $M_{V,host}= -23.2$ $\pm$ 0.2 mag (Misgeld et
al.~\cite{Misgel08}) for NGC 4696. Figure~\ref{distribution} shows
that in Centaurus we detected 18-19 UCDs with $M_V<-11$ mag within the
central $\sim$50 kpc. In the UCD survey from Mieske et
al. (\cite{Mieske07}), within which all but one UCD considered here
were detected, we estimate the survey completeness in terms of slit
allocation and spectroscopic success to be 0.29. However, the region
under consideration in Fig.3 was covered twice in that survey, such
that the completeness improves to 0.29 + (0.29*(1-0.29))=0.50, with an
estimated uncertainty of 0.07. The absolute magnitude $M_{V,host}=
-23.2$ $\pm$ 0.2 mag of NGC 4696 is taken from a FORS2 imaging survey
(Misgeld et al. ~\cite{Misgel08}) which covers the approximately same
pointing, such that we do not have to apply any further corrections in
terms of area completeness to calculate the specific
frequency. The completeness-corrected number of UCDs with $M_V<-11$
mag is therefore 37 $\pm$ 10 in the central Centaurus region, with a
corresponding reference specific frequency of $S_{N,UCD}^* \simeq 13.5
\pm 4.5$, and a final corrected value of $S_{N,UCD}= 5.1 \pm 1.7$. For
comparison, the GC specific frequency of NGC 4696 is $S_{N,GC} = 7.3
\pm 1.5$ (Mieske et al.~\cite{Mieske05}), fully consistent with the
value for UCDs.

\subsection{Hydra}

For the Hydra cluster we base our estimate of $S_{N,UCD}$ on the
recent VIMOS spectroscopic study presented in Misgeld et
al. (\cite{Misgel11}). The assumed distance modulus of Hydra is 33.37
mag (Misgeld et al~\cite{Misgel11}). We restrict our considerations to
a radius of r=5$'$ from the cluster center (60 kpc), adopted to be at
the position of the central cD galaxy NGC 3311. In Misgeld et
al. (\cite{Misgel11}), we found 33 UCDs with $M_V<-11$ mag in this
region. The projected positions of those UCDs are indicated in
Fig.~\ref{distribution}, along with those for Fornax and
Centaurus. The overall completeness in terms of slit allocation is
63\% for this survey. Due to the superposition of several survey
pointings towards the central Hydra cluster, the slit allocation
completeness in the central 5 arcmin is slightly higher, at 70 $\pm$ 5
$\%$. The area coverage within the central 5 arcmin is almost
complete, at about 95 $\%$ (Figs. 1 and 7 of Misgeld et
al.~\cite{Misgel11}). Together, this implies an overall survey
completeness of $\sim$66\%, and hence gives a total number of $\sim
50$ $\pm$ 10 UCDs. For calculating the specific frequency, we adopt an
apparent magnitude of V=10.9 mag for NGC 3311 and V=11.8 for NGC 3309
(Misgeld et al.~\cite{Misgel09}), the two main galaxies in the central
5 arcmin of Hydra. Using as distance modulus 33.37 (Misgeld et
al.~\cite{Misgel11}), we find $M_V=-22.9$ mag as total luminosity of
the two galaxies. This yields a reference specific UCD frequency of
$S_{N,UCD}^* \simeq 22 \pm 8$ for the Hydra cluster. Assuming a GCLF
width of $\sigma=1.30$ mag (corresponding to the $M_V=-22.5$ mag of
the dominant gE NGC 3311), we obtain a final corrected value of
$S_{N,UCD}= 12.5 \pm 4.3$. For comparison, the GC specific frequency
of NGC 3311 within the inner 40 kpc is $S_{N,GC} = 12.5 \pm 1.5$
(Wehner et al.~\cite{Wehner08}), which agrees very well with the UCD
value.

\subsection{Group environments}

\subsubsection{Local Group}
In the Milky Way there is no compact stellar system with $M_V<-11$
mag, and also for M31 it is at most two to three objects (Harris et
al.~\cite{Harris96}, Barmby et al.~\cite{Barmby00}, Huxor et
al.~\cite{Huxor11}).\footnote{M32 is excluded from this consideration}
We therefore use the equivalent definition of $S_{N,UCD}$ based on the
limiting mass of $2 \times 10^6$ M$_{\odot}$ ($M_V=-$10.25 mag), with
$S_{N,UCD} = N_{UCD} 10^{(0.4 M_{V,host}+20)} c_w$. According to
Harris et al. (\cite{Harris96}) there is one compact stellar system in
the Milky Way with $M_V<-10.25$ mag, namely $\omega$Cen. For M31, the
photometry of GCs is affected by both internal and foreground
reddening, which makes the assessment more difficult (e.g. Galleti et
al.~\cite{Gallet04}, Huxor et al.~\cite{Huxor11}). Huxor et
al. (\cite{Huxor11}) find an absolute magnitude distribution for M31
GCs which peaks at $M_V=-7.9$ mag, so is 0.5 mag brighter than the
typical turnover magnitude. Their resulting GC luminosity distribution
(Fig. 2 of their paper), suggests about $\sim$15 GCs in M31 with $M_V
< -10.25$ mag.  However, these authors indicate that uncertainties in
internal extinction can heavily influence those results. Given these
uncertainties, a conservative lower limit on the number of GCs in M31
may be obtained by shifting the overall magnitude distribution of M31
GCs 0.5 mag faintwards to match the typical turnover magnitude (and
that of the Milky Way) of $M_V \sim -7.4$ mag. Doing so, the number of
compact stellar systems with $M_V < -10.25$ mag reduces to $\sim$7-8.

As a sanity check of these numbers, we furthermore consider the NIR
Ks-band luminosities of the brightest M31 GCs from the revised Bologna
Catalogue by Galleti et al.~(\cite{Gallet04}), which are largely
unaffected by reddening. Adopting a typical (V-K)=2.5 mag for GCs
(Bruzual \& Charlot~\cite{Bruzua03} assuming 10 Gyrs and [Fe/H]=-1
dex) and a distance modulus of 24.47 mag to M31 (McConnachie et
al.~\cite{Mcconn05}), we obtain an apparent limiting magnitude of $m_K
\lesssim 11.7$ mag for UCDs in M31. In the revised Bologna Catalogue
we find ten objects with $m_K < 11.7$ mag. This is within the range of
the numbers determined from optical data above.

We thus adopt a final value of ten UCDs with $M_V < -10.25$ mag in
M31, and one UCD in the Milky Way. We adopt as absolute visual magnitude
for M31 $M_V=-21$ mag (Gil de Paz. ~\cite{Gildep07}). For the Milky
Way, the absolute visual magnitude is quite uncertain as a stand-alone
quantity. We adopt a somewhat fainter value of $M_V=-20.5$ mag
compared to M31, based on the approximate mass ratio between both
galaxies when using the radial velocity distribution of their
satellites as dynamical tracers (Watkins et al.~\cite{Watkins10}). The
total absolute magnitude for the Local Group then is $M_V=-21.5$
mag. For the Local Group we thus obtain a reference specific frequency
of $S_{N,UCD}^*=2.8$ $\pm$ 1.0. Evaluating $c_w$ for $M_V=-21$ mag
(since M31 contributes almost all of the UCDs), we get $\sigma=1.15$
mag and a final corrected value of $S_{N,UCD}=4.1$ $\pm 1.5$. For
  comparison, the GC specific frequency of M31 is
  $S_{N,GC} = 1.8 \pm 0.3$ (Barmby \& Huchra~\cite{Barmby01} \& Gil de
  Paz et al.~\cite{Gildep07}), marginally lower than the UCD specific
  frequency at the 1.5 $\sigma$ level. 

\subsubsection{Hickson compact groups 22 and 90}

In a recent study by Da Rocha et al. (\cite{Daroch11}), a
spectroscopic search for UCDs is performed in HCG 22 and 90 (distance
modulus 32.6 mag) using FORS2 MXU spectroscopy in one 7x7$'$ pointing
(65x65 kpc). Also in this environment, no bright UCDs with $M_V<-11$
mag have been found, so we adopt the same definition of $S_{N,UCD}$ as
for the Local Group. In Da Rocha et al. (\cite{Daroch11}) the specific
frequency of UCDs with mass above $2 \times 10^6$ M$_{\odot}$ is
already calculated taking the survey completeness
(Sects. 2.2 and 4.2. of that paper) into account; however, in this paper
the authors adopt a definition based on the B-band luminosity of the
host galaxy: $S_{N,UCD}^* = N_{UCD} 10^{(0.4 (M_{B,host}+20)}$. For
this normalization, the specific frequency values are 6.3 $\pm$ 2.1
for HCG 22 and 2.0 $\pm$ 1.0 for HCG 90, for total B-band group
luminosities of $M_B=-20.7$ and $M_B=-21.5$, respectively. For HCG 90
this includes a contribution of $\sim 35\%$ intragroup light (Da Rocha
et al. \cite{Daroch11}). To convert this to the definition based on
V-band presented in this paper, we assume a global (B-V)=0.9 mag for
both group environments. This scales down the numerical values by a
factor $10^{0.9 \times 0.4}$ and thus yields reference specific
frequencies of $S_{N,UCD}^*=$2.7 $\pm$ 1.2 and $S_{N,UCD}^*=$0.9 $\pm$
0.5 for HCG 22 and HCG 90. Including the correction factor $c_w$ with
$\sigma=1.15$ mag in both groups (main galaxy luminosities of
$M_V=-21.1$ mag), we obtain final values of $S_{N,UCD}=$3.9 $\pm$ 1.6
and $S_{N,UCD}=$1.3 $\pm$ 0.7 for HCG 22 and HCG 90,
respectively. The value for HCG 90 would increase to $S_{N,UCD}
  \sim 2 \pm 1$ when excluding the intragroup light component from the
  total group luminosity.

\begin{figure}
\begin{center}
  \includegraphics[width=8.6cm]{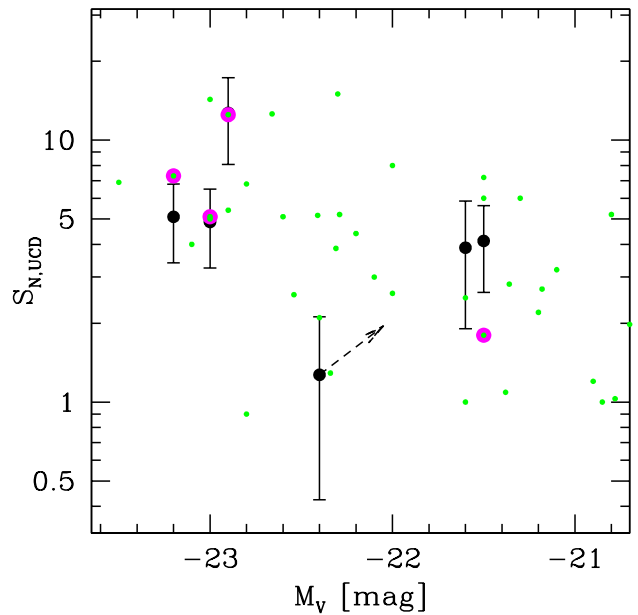}
  \caption{Specific frequency of UCDs. The large filled
    circles with error bars indicate the UCD specific frequency
    $S_{N,UCD}$ as derived in the text for six different
    environments. These are, from bright to faint host galaxy
    luminosities: Centaurus, Fornax, Hydra, HCG90, HCG22, Local
    Group. See also Table~\ref{tablesn}. Large magenta circles are the GC
    specific frequencies $S_{N,GC}$ in the investigated environments, see text for details -- estimates for the HCGs are not available. Small (green) dots
    are literature GC specific frequencies $S_{N,GC}$
    over a range of environments (Dirsch et al. (\cite{Dirsch03}),
    Mieske et al. (\cite{Mieske05}), Peng et al. (\cite{Peng08}) and
    references therein). The dashed arrow for HCG90 indicates the location of this data point if the 35\% intragroup light component (see text) would not be considered for the total luminosity.}
\label{sn}
\end{center}
\end{figure}

\subsection{Global picture of UCD specific frequencies}

The calculations performed in the previous sections regarding specific
frequencies are summarised in Table~\ref{tablesn} and in
Fig.~\ref{sn}, which shows the UCD specific frequency $S_{N,UCD}$ as
derived for the six different environments. For comparison, it also
shows the GC specific frequency $S_{N,GC}$ in the respective
environments. Data on $S_{N,GC}$ for the HCGs are not available. We
also indicate literature globular cluster
specific frequencies $S_{N,GC}$ (Dirsch et al.~\cite{Dirsch03}, Mieske
et al.~\cite{Mieske05}, Peng et al.~\cite{Peng08} and references
therein) over a range of environments. The GC specific frequencies
rise slightly for the brightest host galaxy magnitudes $M_V \lesssim
-23$ mag, which has been noted in a number of previous studies
(e.g. Kundu \& Whitmore~\cite{Kundu01}, Brodie \&
Strader~\cite{Brodie06} and references therein, Jord\'{a}n et
al.~\cite{Jordan06}, Peng et al.~\cite{Peng08}, Georgiev et
al.~\cite{Georgi10}).

Overall, the specific frequencies of UCDs match those
of GCs remarkably well. For the four environments where both $S_{N,UCD}$ and
$S_{N,GC}$ are available, we find an average of $S_{N,UCD}=6.7 \pm$
2.0 vs. $S_{N,GC}=6.7 \pm 2.2$. The mean over all six environments is
$S_{N,UCD}=$5.3 $\pm$ 1.7, fully consistent with the mean
$S_{N,GC}=$5.0 $\pm$ 0.7 averaged over all galaxies in the host
magnitude range $M_V<-21$ mag. We find mild evidence that UCDs have a
higher specific frequency in more luminous hosts, as also seen for GCs
(e.g. Peng et al.~\cite{Peng08}).  Adopting a dividing host galaxy
luminosity at $M_V=-22.5$ mag (galaxy cluster vs. galaxy group), we
find for UCDs that $S_{N,UCD,bright} = 7.6 \pm 2.6$ while
$S_{N,UCD,faint} = 3.1 \pm 0.9$. This difference of about a factor of
2-3 is formally different from zero at the 1.6 $\sigma$ level. For GCs
we find $S_{N,GC,bright} = 6.4 \pm 1.4$ and $S_{N,GC,faint} = 4.3 \pm
0.8$.

We note that UCDs have also been detected in the Virgo and Coma
clusters (e.g. Jones et al.~\cite{Jones06}, Chiboucas et
al.~\cite{Chibou10a}), but the respective survey designs do not allow
accurate determination of the UCD specific frequency. To solidify
the suggested increase in UCD frequency in high-luminosity and denser
environments, the results of homogeneous spectroscopic surveys down to
$M_V \simeq -10$ to $-11$ mag in both Virgo and Coma will be very
helpful.

\begin{table*}
\caption{List of the specific frequencies $S_{N,UCD}$ derived in
  Sect.~\ref{envsn}, along with the total absolute magnitude of the
  host environments. $M_{V,host}$ includes the full galaxy light in
  the UCD survey area (see text). $M_{V,host,individual}$ gives the magnitude of
  the brightest galaxy in the environment, which defines the assumed
  width $\sigma$ of the GCLF and hence the correction term $c_w$. The
  last column gives the literature value for the GC specific frequency
  of the brightest cluster galaxy, where available. $^*$ Mieske et
  al. ~\cite{Mieske05}; $^{**}$ Dirsch et al. ~\cite{Dirsch03};
  $^{***}$ Wehner et al. ~\cite{Wehner08}; $^{****}$ Barmby \&
  Huchra~\cite{Barmby01} \& Gil de Paz. ~\cite{Gildep07}.}
\begin{center}
\begin{tabular}{|lrrrrl|}\hline
Environment & $S_{N,UCD}$ & $M_{V,host}$ & $M_{V,host,individual}$& $c_w$ & $S_{N,GC}$  \\\hline
Centaurus &5.1 $\pm$ 1.7& -23.2 & -23.2 & 0.38 & 7.3 $\pm$ 1.5 $^*$\\
Fornax & 4.9 $\pm$ 1.7& -23.0 & -22.6 & 0.54 & 5.1 $\pm$ 1.2$^{**}$\\
Hydra & 12.5 $\pm$ 4.3& -22.9 & -22.5& 0.58 & 12.5 $\pm$ 1.5$^{***}$\\
HCG 90 &1.3 $\pm$ 0.7&-22.4 & -21.1 & 1.41 &\\
HCG 22 & 3.9 $\pm$ 1.6&-21.6 & -21.1 & 1.41 &\\
Local Group &4.1 $\pm$ 1.5&-21.5& -21.0 &1.5& 1.8 $\pm$ 0.3$^{****}$\\\hline
\end{tabular}
\end{center}
\label{tablesn}
\end{table*}

\section{Number of UCDs compared to number of possible progenitor galaxies}
\label{progen}

As outlined in the Introduction, there are two main formation channels
for UCDs discussed in the literature, that of massive star clusters
and that of tidally stripped dwarf galaxies
(e.g. Zinnecker~\cite{Zinnec88}, Hilker et al.~\cite{Hilker99b},
Drinkwater et al.~\cite{Drinkw00}, Bekki et al.~\cite{Bekki03}, Goerdt
et al.~\cite{Goerdt08}). In the following we discuss the latter
scenario, and use the result of the previous section to constrain its
importance. The general idea is that an overabundance of UCDs with
respect to GCs would indicate there is a second formation channel. An
upper limit for such an overabundance of UCDs with respect to GCs can
be estimated from the ratio of average UCD-to-GC specific frequencies
and its error bars: $\frac{6.7 \pm 2.0}{6.7 \pm 2.2}=1.00 \pm
0.44$. The error bars of this ratio suggest that no more than 50\% of
UCDs may be formed via tidally stripped dwarf galaxies (2 $\sigma$
limit).

It is worth noting that the luminosity function of dwarf galaxy nuclei
is offset by two to three mag brighter with respect to their GC
systems (Lotz et al.~\cite{Lotz01}, Mieske et
al. ~\cite{Mieske04}). Assuming a Gaussian LF, an additional $\sim$1\%
of objects with a LF peaking two to three mag brighter than average
GCs double the number counts in the UCD luminosity regime.

In the following we compare the number of UCDs in
Hydra, Centaurus, Fornax to the number of known dwarf galaxies in the
same area whose luminosities are in the range expected for UCD
progenitors. The null hypothesis is that all UCDs are formed via tidal
threshing of nucleated dwarf galaxies (Bekki et
al.~\cite{Bekki03}). Under this hypothesis we assess what
fraction of a primordial dwarf galaxy population would need to have
already been disrupted to account for all UCDs.

Bekki et al. (\cite{Bekki03}) perform simulations of tidal stripping
of nucleated dwarf galaxies to investigate this formation channel for
UCDs. They adopt a mass fraction of a few percent of the central
nucleus compared to the stellar envelope of the host dwarf galaxy (a
difference of 3-4 mag). The recent observational studies performed in
the course of the ACS Virgo Cluster Survey (e.g. C\^{o}t\'{e} et
al.~\cite{Cote06}) suggest, however, that the mass fraction of nuclei
is one order of magnitude less, about 0.3\% or 6-7 mag.

To estimate the luminosity range for possible UCD progenitors, we
first define as starting point the luminosity range of UCDs themselves
to be $-13.5<M_V<-11$ mag. For the UCD progenitors, we then adopt a
magnitude difference of 6 $\pm$ 1 mag, which yields a magnitude
range $-20.5 < M_V < -16$ for potential UCD progenitor galaxies.

The left panel of Fig.~\ref{distribution} shows the map of the
projected positions of UCDs in the central $50$kpc of the Hydra,
Fornax and Centaurus clusters, adopting as UCD limit $M_V<-11$ mag
(see previous section). In the right panel of Fig.~\ref{distribution}
we show in addition the projected positions of the potential UCD
progenitor galaxies, that is, those galaxies with $-20.5 < M_V < -16$
mag. For Hydra, the galaxy positions and magnitudes are taken from a
merging of the FORS2 Hydra Cluster Catalog (Misgeld et
al.~\cite{Misgel09}) and the spectroscopic study of Christlein \&
Zabludoff (\cite{Christ03}). For Centaurus, the galaxies are from a
merging of the Centaurus Cluster Catalog (Stein et al.~\cite{Stein97})
and the FORS2 photometric study of Misgeld et
al. (\cite{Misgel08}). For the Fornax cluster, the galaxies are from a
merging of the Fornax Cluster Catalogue (Ferguson \&
Sandage~\cite{Fergus88}) and the photometric study of Mieske et
al. (\cite{Mieske07}) performed with IMACS@Magellan.

For the right panel of Fig.~\ref{distribution} we assume that the
galaxy catalogs for cluster members in the three environments are
complete within the magnitude range considered. The UCD searches in
the same area have a completeness significantly below 100\%
(see previous section). Averaging over the three clusters and
weighting with the respective number of detected UCDs, we obtain an
estimate of $\sim 62 \pm 5 \%$ for the completeness of the UCD
search\footnote{$(0.7 12 + 0.5 18.5 + 0.66 33)/63.5=0.62$}. In the
right panel of Fig.~\ref{distribution} we therefore plot only 62\% of
all eligible progenitor galaxies in this figure, which allows a direct
comparison of the number counts. We furthermore highlight the very
brightest subsample of UCDs, adopting a (somewhat arbitrary) limit of
$M_V<-12$ mag.

This plot shows that, within the same projected radius, the number of
UCDs with $M_V<-11$ mag is a factor of 6-7 larger than the number of
existing possible progenitor galaxies. Under the null hypothesis that
all UCDs are created by tidal stripping, this plot suggests that at
least $\frac{54}{54+9} \simeq 85 \%$ of primordial dwarf galaxies in
the central $\sim$50-70 kpc of the considered galaxy clusters would
have had to be tidally disrupted already. The situation is naturally
less extreme when restricting to the very brightest UCDs with $M_V
<-12$ mag. For such a scenario one would expect a more modest two thirds of primordial dwarfs to have been tidally disrupted by
now.

In the right panel Fig.~\ref{fornax_rad}, we present a more extended
look at the situation in the Fornax cluster, given the large available
UCD database. We show a comparison of the cumulative radial
distribution of UCDs and progenitor dwarf galaxies out to 200 kpc
clustercentric distance. The curves are based on
Fig.~\ref{fornax_map}, and have been renormalized to represent the
relative size of each subsample. It is again very clear from this plot
that UCDs vastly outnumber present-day progenitor galaxies. Compared
to the full UCD sample down to $M_V<-10.25$ mag, the number of
progenitor dwarfs is only a few percent. UCDs are much more centrally
clustered. If indeed a significant amount of UCDs were formed from
tidal processes, then the tidal disruption of low-mass dark matter
halos in the central Fornax cluster must have been extremely efficient
($\gtrsim 90\%$).

We note that the particular choice of 6 mag for the magnitude
difference between progenitor and UCD influences the considerations
above to some extent, since the galaxy LF increases
towards fainter luminosities. A smaller magnitude difference between
UCDs and progenitors will slightly increase the number of eligible
progenitor galaxies. When assuming a mean magnitude difference of 4
mag instead of 6 mag between UCDs and progenitors, the number of
progenitors changes from 9 to 12 within 50 kpc for the joint
Hydra-Fornax-Centaurus sample. Still, this is far from the number of
existing UCDs.

Bekki et al. (\cite{Bekki03}) performed a theoretical study tailored
precisely to the case of UCD formation via tidal stripping of dwarf
galaxies. They determine a ``threshing radius'', postulating that
dwarf galaxies with orbital pericenters within that radius would
become tidally transformed to UCDs. This concept is also employed by
the analytical study in Thomas et al.~(\cite{Thomas08}). For typical
dwarf galaxy luminosities/masses, Bekki et al. predict the threshing
radius to be in the range 50-100 kpc for the Fornax cluster, and
70-150 kpc for the Virgo cluster (Fig. 7 of their paper).  Therefore,
the strong central clustering of UCDs (Fig.~\ref{fornax_map}) and the
implied scenario where $\gtrsim$ 90\% of primordial dwarf galaxies in
the central $\sim$50-70 kpc are disrupted, are in qualitative
agreement with this model.

There are some more recent studies in the literature considering the
tidal disruption of satellite galaxies (e.g. Henriques et
al.~\cite{Henriq08}, Henriques \& Thomas~\cite{Henriq10}, Yang et
al.~\cite{Yang09}) and their contribution to the intracluster
light. However, those studies were not specifically tailored to
address the tidal stripping scenario for UCD formation. 

Yang et al. (\cite{Yang09}) argue that tidal disruption of satellites
will be more efficient for {\it high} ratios between satellite and
host halo mass. At the same time, they show that the stellar mass
contributed to the intracluster medium by dissolving satellites is
higher for massive host halos $\sim 10^{14}$ M$_{\odot}$ than for
lower mass halos $\sim 10^{12}$ M$_{\odot}$. 

Henriques et al. (\cite{Henriq08}) estimate that about one half of
satellite galaxies get disrupted and/or accreted to their host halos
and that 10\% of the overall cluster light is found in the
intracluster medium, originating from tidally stripped stars. In their
most recent paper, Henriques \& Thomas (\cite{Henriq10}) improve their
treatment of tidal disruption. In turn they revise the intra-cluster
light fraction provided by disrupted galaxies upwards to 30\%.  Their
models predict 40\% of the satellites with masses between $10 ^9$ and
$10 ^{10}$ $M_{\odot}$ to be tidally disrupted. This fraction
increases to 57\% when considering satellites that have already lost
their dark matter halo (``orphan satellites'') in previous
interactions (Henriques 2011, private communication).

This $\sim$50\% global disruption fraction is lower than the fraction
of $\sim$90\% needed for the limiting case that half of today's UCDs
originate from tidal processes (see above). Radial trends of
disruption efficiency would still be expected (Bekki et
al.~\cite{Bekki03}), such that this disruption fraction can be
considered a lower limit to the value applicable to the clustercentric
areas studied in this paper.

\section{Summary and conclusions}
\label{conclusions}

In this paper we have proposed a definition for the specific frequency
of UCDs, which we denote as $S_{N,UCD}$. We adopt the following
functional form:\\

$S_{N,UCD} = N_{UCD}   10^{0.4    (M_{V,host}-M_{V,0})}   c_{w}$.\\

This definition normalizes the number of UCDs in a given environment
to a unit host galaxy luminosity $M_{V,host}$, analogous to the
definition of the specific frequency $S_{N,GC}$ for GCs. The premise
of our definition is that if UCDs follow the extrapolation of the GCLF
to bright magnitudes, then $S_{N,UCD} = S_{N,GC}$. This premise
defines the value of the zero point $M_{V,0}$. Considering UCDs as
compact stellar systems with $M_V<-10.25$ mag, we find that the value
of $M_{V,0}=-20$ mag fulfills the above premise. For the case of
extragalactic surveys with a brighter completeness limit $M_V<-11.0$
mag, we need to adopt $M_{V,0}=-22$ mag. The term $c_{w}$ is
introduced to correct the specific frequency for the well known
systematic variation of the GCLF width $\sigma$ with host galaxy
magnitude. Details of this correction are outlined in
Sect.~\ref{sndefsec}. In Sect.~\ref{envsn} we apply our proposed
definition of $S_{N,UCD}$ to results of spectroscopic UCD searches
performed by our group in the Fornax, Hydra and Centaurus galaxy
clusters, and two Hickson Compact Groups. We also include the Local
Group.

Our main finding is that the specific frequencies derived for UCDs
match those of GCs very well. For four of the six investigated
environments, there are GC specific frequency measurements available,
allowing a direct comparison. For those four environments we find a
mean $S_{N,UCD}=6.7 \pm$ 2.0, vs. $S_{N,GC}=6.7 \pm 2.2$. The ratio of
UCD-to-GC specific frequency is therefore $\frac{6.7 \pm 2.0}{6.7 \pm
  2.2}=1.00 \pm 0.44$. The mean $S_{N,UCD}$ of all six investigated
environments is 5.3 $\pm$ 1.7, in agreement with the average GC
specific frequency of 5.0 $\pm$ 0.7 of the available literature data
for the corresponding host magnitude range $M_V<-21$ mag. Our findings
are consistent with the hypothesis that most UCDs are formed by the
same process as the overall GC population, and with a similar
formation efficiency to GCs.

 We also present an extension of our analysis in the Fornax cluster by
 using the large available data set of spectroscopically confirmed UCDs
 and GCs. This literature data set comprises about 180 confirmed UCDs
 with $M_V<-10.25$ mag. We find that the specific frequencies of UCDs
 around giant galaxies and in the intracluster space are consistent
 with their being drawn from the bright tail of the GCLF. There is
 still room for a possible UCD overabundance in the intracluster
 space by a factor of $\sim 2$, given the incompleteness in
 spectropscopic coverage in those regions. We do not find significant
 evidence of a different spatial distribution between UCDs and GCs.

It has been proposed that the present-day population of UCDs is indeed
{\it a superposition} of sources formed via tidal stripping of dwarf
galaxies and sources formed in the same process as the main GC
population (e.g. Ha\c{s}egan et al.~\cite{Hasega05}, Mieske et
al.~\cite{Mieske06}, Chilingarian et al.~\cite{Chilin11}, Norris et
al.~\cite{Norris11}, da Rocha et al.~\cite{Daroch11}). The error bars
of the specific frequencies derived for UCDs suggest $\sim$50 \% as an
upper limit for the importance of the dwarf galaxy channel. We show in
Sect.~\ref{progen} that this would require at least $\gtrsim$ 90\% of
primordial dwarf galaxies in the central $\sim$50-70 kpc of the
considered galaxy clusters to have already been disrupted. If indeed a
significant amount of UCDs were formed from tidal processes, then the
tidal stripping of stars from low-mass dark matter halos in the
central Fornax cluster has been extremely efficient.\\

\noindent
We conclude that the number counts of UCDs are fully consistent with
them being the bright tail of the GC population. From a
statistical point of view there is no need to invoke an additional
formation channel. The statistical error bars constrain the fraction
of tidally stripped dwarfs to not more than 50\% of UCDs.

\acknowledgements We thank the anonymous referee for very useful
comments and suggestions that helped to improve the paper. We thank
Bruno Henriques for providing us with details on the number fraction
of disrupted satellite galaxies in their models. IM acknowledges
support through DFG grant BE1091/13-1.

\end{document}